\documentclass[apj]{emulateapj}
\usepackage{epsfig}
\usepackage{epstopdf}
\usepackage{graphics}

\begin{document}

%=====TITLE AND ABSTRACT=======================================================

\title[]{Northern Sky Galactic Cosmic Ray Anisotropy between 10 and 1000 TeV with the Tibet Air Shower Array}
\author{
M.~Amenomori$^{1}$, X.~J.~Bi$^{2}$, D.~Chen$^{3}$,
T.~L.~Chen$^{4}$, W.~Y.~Chen$^{2}$, S.~W.~Cui$^{5}$,
Danzengluobu$^{4}$, L.~K.~Ding$^{2}$, C.~F.~Feng$^{6}$,
Zhaoyang~Feng$^{2}$, Z.~Y.~Feng$^{7}$, Q.~B.~Gou$^{2}$,
Y.~Q.~Guo$^{2}$, H.~H.~He$^{2}$, Z.~T.~He$^{5}$,
K.~Hibino$^{8}$, N.~Hotta$^{9}$, Haibing~Hu$^{4}$,
H.~B.~Hu$^{2}$, J.~Huang$^{2}$, H.~Y.~Jia$^{7}$,
L.~Jiang$^{2}$, F.~Kajino$^{10}$, K.~Kasahara$^{11}$,
Y.~Katayose$^{12}$, C.~Kato$^{13}$, K.~Kawata$^{14}$,
M.~Kozai$^{13,a}$, Labaciren$^{4}$, G.~M.~Le$^{15}$,
A.~F.~Li$^{16,6,2}$, H.~J.~Li$^{4}$, W.~J.~Li$^{2,7}$,
C.~Liu$^{2}$, J.~S.~Liu$^{2}$, M.~Y.~Liu$^{4}$,
H.~Lu$^{2}$, X.~R.~Meng$^{4}$, T.~Miyazaki$^{13}$,
K.~Mizutani$^{11,17}$, K.~Munakata$^{13}$, T.~Nakajima$^{13}$,
Y.~Nakamura$^{13}$, H.~Nanjo$^{1}$, M.~Nishizawa$^{18}$,
T.~Niwa$^{13}$, M.~Ohnishi$^{14}$, I.~Ohta$^{19}$,
S.~Ozawa$^{11}$, X.~L.~Qian$^{6,2}$, X.~B.~Qu$^{2}$,
T.~Saito$^{20}$, T.~Y.~Saito$^{21}$, M.~Sakata$^{10}$,
T.~K.~Sako$^{14}$, J.~Shao$^{2,6}$, M.~Shibata$^{12}$,
A.~Shiomi$^{22}$, T.~Shirai$^{8}$, H.~Sugimoto$^{23}$,
M.~Takita$^{14}$, Y.~H.~Tan$^{2}$, N.~Tateyama$^{8}$,
S.~Torii$^{11}$, H.~Tsuchiya$^{24}$, S.~Udo$^{8}$,
H.~Wang$^{2}$, H.~R.~Wu$^{2}$, L.~Xue$^{6}$,
Y.~Yamamoto$^{10}$, K.~Yamauchi$^{12}$, Z.~Yang$^{2}$,
%cS.~Yasue$^{25}$,
 A.~F.~Yuan$^{4}$, T.~Yuda$^{14,b}$,
L.~M.~Zhai$^{3}$, H.~M.~Zhang$^{2}$, J.~L.~Zhang$^{2}$,
X.~Y.~Zhang$^{6}$, Y.~Zhang$^{2}$, Yi~Zhang$^{2}$,
Ying~Zhang$^{2}$, Zhaxisangzhu$^{4}$, X.~X.~Zhou$^{7}$\\
(The Tibet AS$\gamma$ Collaboration)\\
}
\affil{
{$^{1}$}{Department of Physics, Hirosaki University, Hirosaki 036-8561, Japan}\\
{$^{2}$}{Key Laboratory of Particle Astrophysics, Institute of High Energy Physics, Chinese Academy of Sciences, Beijing 100049, China}\\
{$^{3}$}{National Astronomical Observatories, Chinese Academy of Sciences, Beijing 100012, China}\\
{$^{4}$}{Department of Mathematics and Physics, Tibet University, Lhasa 850000, China}\\
{$^{5}$}{Department of Physics, Hebei Normal University, Shijiazhuang 050016, China}\\
{$^{6}$}{Department of Physics, Shandong University, Jinan 250100, China}\\
{$^{7}$}{Institute of Modern Physics, Southwest Jiaotong University, Chengdu 610031, China}\\
{$^{8}$}{Faculty of Engineering, Kanagawa University, Yokohama 221-8686, Japan}\\
{$^{9}$}{Faculty of Education, Utsunomiya University, Utsunomiya 321-8505, Japan}\\
{$^{10}$}{Department of Physics, Konan University, Kobe 658-8501, Japan}\\
{$^{11}$}{Research Institute for Science and Engineering, Waseda University, Tokyo 169-8555, Japan}\\
{$^{12}$}{Faculty of Engineering, Yokohama National University, Yokohama 240-8501, Japan}\\
{$^{13}$}{Department of Physics, Shinshu University, Matsumoto 390-8621, Japan}\\
{$^{14}$}{Institute for Cosmic Ray Research, University of Tokyo, Kashiwa 277-8582, Japan}\\
{$^{15}$}{National Center for Space Weather, China Meteorological Administration, Beijing 100081, China}\\
{$^{16}$}{School of Information Science and Engineering, Shandong Agriculture University, Taian 271018, China}\\
{$^{17}$}{Saitama University, Saitama 338-8570, Japan}\\
{$^{18}$}{National Institute of Informatics, Tokyo 101-8430, Japan}\\
{$^{19}$}{Sakushin Gakuin University, Utsunomiya 321-3295, Japan}\\
{$^{20}$}{Tokyo Metropolitan College of Industrial Technology, Tokyo 116-8523, Japan}\\
{$^{21}$}{Max-Planck-Institut f\"ur Physik, M\"unchen D-80805, Deutschland}\\
{$^{22}$}{College of Industrial Technology, Nihon University, Narashino 275-8576, Japan}\\
{$^{23}$}{Shonan Institute of Technology, Fujisawa 251-8511, Japan}\\
{$^{24}$}{Japan Atomic Energy Agency, Tokai-mura 319-1195, Japan}\\
{$^{a}$ }{now at: ISAS/JAXA Sagamihara 252-5210, Japan}\\
{$^{b}$ }{Deceased.}\\
%c\affil{$^{25}$}{School of General Education, Shinshu University, Matsumoto 390-8621, Japan}\\
}

\begin{abstract}

We report on the analysis of the $10-1000$ TeV large-scale sidereal
anisotropy of Galactic cosmic rays (GCRs) with the data collected by
the Tibet Air Shower Array from 1995 October to 2010 February.
In this analysis, we improve the energy estimate and
extend the decl. range down to $-30^{\circ}$. We find that the
anisotropy maps above 100 TeV are distinct from that at a multi-TeV
band. The so-called tail-in and loss-cone features
identified at low energies get less significant, and a new component
appears at $\sim100$ TeV. The spatial distribution of the GCR intensity with an excess (7.2$\sigma$ pre-trial,
5.2$\sigma$ post-trial) and a deficit ($-5.8\sigma$ pre-trial) are
observed in the 300 TeV anisotropy map, in close agreement with IceCube's results at 400 TeV. Combining the Tibet
results in the northern sky with IceCube's results in the southern
sky, we establish a full-sky picture of the anisotropy in
hundreds of TeV band. We further find that the amplitude of the first order
anisotropy increases sharply above $\sim100$ TeV, indicating a new
component of the anisotropy. All these results may shed new light on
understanding the origin and propagation of GCRs.

\end{abstract}

\maketitle

\section{Introduction}
The arrival directions of Galactic cosmic rays (GCRs) are nearly
isotropic due to deflections in the Galactic magnetic field (GMF).%c \citep{2005ApJ...626L..29A}
 Only weak anisotropy is expected from the diffusion and/or
 drift of GCRs in GMF. Observations of ground-based air shower
arrays and underground muon detectors do show the existence of small
anisotropies with relative amplitudes of the order of
$10^{-4}$ to $10^{-3}$ at energies from 100 GeV
to hundreds of TeV (see Figure~\ref{fig1DFit}).
 However, the
variation of the amplitude with energy seems to be difficult to interpret in terms of
 the conventional GCR diffusion model in the Galaxy ($e.g.$ \citep{2002ApJ...565..280M,2016arXiv161201873A}).
 The study of GCR anisotropy, therefore, is important to understand the origin and propagation of GCRs.

Only a few results of the anisotropy in the energy range from hundreds
of TeV up to $\sim10$ PeV have been reported, primarily due to the
low fluxes of cosmic rays (CRs) in this energy range. EAS-TOP
collaboration reported for the first time a detection of anisotropy at
$\sim200$~TeV \citep{1996ApJ...470..501A}. With the accumulation of
data, they improved their result later and reported a sharp increase
of the anisotropy amplitude at primary energies around $\sim$370~TeV
\citep{2009ApJ...692L.130A}. At the PeV energy region, the Akeno experiment  reported an increase of the CR anisotropy amplitude in 1986 \citep{1986JPhG...12..129K}. No hint of the anisotropy, on the other hand, has been found in the KASCADE data at higher energies between 0.7 and 6 PeV \citep{2004ApJ...604..687A}.
 Recently, the IceCube collaboration
reported the anisotropy observed in the southern sky, showing a new feature different from that obtained by EAS-TOP
\citep{2012ApJ...746...33A}. A clear deficit with a post-trial
significance of $-6.3\sigma$ at 400 TeV was detected, which was then
confirmed by the result from Ice-Top \citep{2013ApJ...765...55A}.
The Ice-Top data further revealed the existence of anisotropy at energies up to 1 PeV \citep{2013ApJ...765...55A}.

The Tibet Air Shower (AS) array
collaboration presented the first two-dimensional anisotropy
measurements in an energy region from several TeV to several hundred TeV.
The anisotropy features, known as the ``tail-in'' and ``loss-cone'' features,
were observed with very high significances  \citep{2006Sci...314..439A}. A new component
anisotropy at multi-TeV energies from the Cygnus direction was also
reported \citep{2006Sci...314..439A}. It has been shown that the amplitude of the first order anisotropy decreases above a few hundred TeV, indicating the
co-rotation of GCRs around the Galactic center. With more data accumulated,
hints of $\sim300$ TeV anisotropies have been revealed
\citep{ZFeng2009,ZFeng2013}. The anisotropy feature was found to
be different from those in lower energy regions and in agreement with
IceCube's result at 400 TeV \citep{2012ApJ...746...33A}. These analyses of Tibet AS array data
 cover decl. from
$-15^{\circ}$ to $75^{\circ}$, yet leaving a gap to be connected with
IceCube's result in the southern sky. Here we extend these analyses to include
events with zenith angle up to $60^{\circ}$, which corresponds to a coverage
of decl. from $-30^{\circ}$ to $90^{\circ}$
\citep{ZFeng2015}. Combining with IceCube's results, we  present for the
first time a full-sky anisotropy observed at hundreds of TeV.
By improving the reconstruction of primary energy, we will also extend the analyzed energy range
 to two decades between 10 TeV and 1 PeV,
which is also the widest coverage of such works.

\section{Analysis}

\subsection{Experiment and Data reconstruction}
The Tibet AS Array is located at Yangbajing in Tibet, China
($90.522^{\circ}$E, $30.102^{\circ}$N, 4300 m above sea level, 606
g/cm$^{2}$ atmospheric depth).
The detector array consists of plastic
scintillation detectors with an area of 0.5 m$^{2}$ each. The
effective area of the Tibet AS array has been gradually enlarged,
via adding the same-type detectors to the array.
 The Tibet I array was constructed in 1990,
 using 65 plastic scintillation detectors placed on grids
 with 15 m spacing. It was then upgraded to 221 detectors on 15 m grids, covering a total of 36,900 m$^{2}$, known as the Tibet II
array. It began operation in 1995 October, with a trigger
rate of $\sim230$ Hz. The Tibet II was then upgraded to the current Tibet III, a
denser array with 7.5 m grids, in 1999 and 2003
\citep{2003ApJ...598..242A}. The trigger rate is $\sim$1700 Hz for
the Tibet III array.

In order to maintain the uniformity of the array performance,  we analyze the data
 keeping the same configuration of the Tibet II array throughout the observation period from 1995 October
to 2010 February, so that the full data sample
taken by Tibet II and Tibet III array can be used in the present
analysis.
%cWe use the full data taken by the Tibet-II and -III arrays in the
%cpresent analysis, which covers the period from October, 1995 to
%cFebruary, 2010.
The traditional shower reconstruction procedure is
applied to get all the parameters of one shower, such as the core
position, zenith, and azimuth angles ($\theta$, $\phi$) of the incident direction and shower size
$\sum\rho_{FT}$ (the sum of the number of particles per m$^{2}$
counted by all the fast-timing [FT] detector). The following three
criteria are applied to select events for further analyses: (1) each AS event should
fire four or more detectors, with each recording 1.25 or more
particles; (2) the AS core position should be located inside the
array; and (3) zenith angle $\theta < 60^{\circ}$.

\subsection{Estimation of the CR Energy}
The ASs reaching the array with a large zenith angle $\theta$  travel through a larger slant atmospheric depth than the vertical ones. This
leads to a zenith angle dependence of the relation between
$\sum\rho_{FT}$ and the primary particle energy. In most of the
previous works of the Tibet AS$\gamma$ Collaboration
\citep{2003ApJ...598..242A,2005ApJ...633.1005A,2006Sci...314..439A,ZFeng2009,ZFeng2013}, the shower size $\sum\rho_{FT}$ is solely
adopted to infer the primary energy of an AS without considering the
zenith angle dependence. This approximation works well for small zenith angles ($\theta<\sim40^{\circ}$), considering the
natural fluctuation in the development of the AS and the limited
resolution of the primary energy.

In this work, we intend to explore the anisotropy with decl.
down to $-30^{\circ}$ by including showers with zenith angles up to
$60^{\circ}$. Therefore, the zenith angle dependence of
the energy reconstruction must be taken into account. We develop a
two-dimensional selection criterion in the $\sum\rho_{FT}-sec\theta$
plane for the energy reconstruction.

The uncertainty of the CR energy reconstruction has been estimated
with a full Monte Carlo (MC) simulation of CR interactions in the
atmosphere by CORSIKA (version 6.204; \citep{1998cmcc.book.....H}.
The hadronic interaction model QGSJET01c and the
detector response modeled by Epics (version 8.65;
\citep{EPICS-website} are used. In this simulation, we adopt the composition
and spectrum models of primary CRs given in
\citep{2003APh....19..193H} as inputs.

\begin{figure}
\centering
\includegraphics[width=7cm, height=4cm]{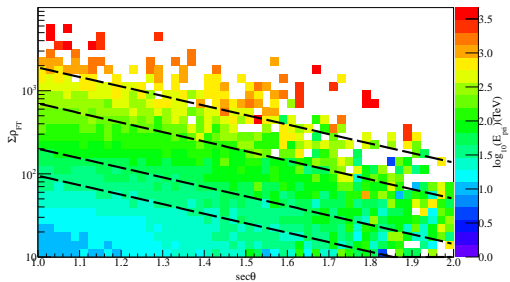}
\caption{\label{figZenithEenrgy} Simulated distribution of the log-mean energy of
 primary CRs as a function of $\sum\rho_{FT}$ and zenith angle.
The $y$-axis is $\sum\rho_{FT}$; the $x$-axis is sec $\theta$, where
$\theta$ is the reconstructed zenith angle; and the color scale
represents the reconstructed log-mean energy in
units of TeV. Dashed lines mark out the borders of events with
different energies (15, 50, 100, 300, and 1000 TeV). }
\end{figure}

Figure~\ref{figZenithEenrgy} shows the simulated distribution in the
primary CR energy as a function of $\sum\rho_{FT}$ and
sec $\theta$. It shows that for a given range of $\sum\rho_{FT}$,
small zenith angle events (at sec $\theta\sim1$) are dominated by
CRs with lower average energy, compared with large zenith angle
events (at sec $\theta\sim2$). We show regions of constant primary
energies (15, 50, 100, 300, and 1000 TeV) in the plane of
($\sum\rho_{FT}$, sec $\theta$) as regions delimited by the dashed lines in
Figure~\ref{figZenithEenrgy}. This grouping enables us to select events in five
 energy samples with minimal overlappings.

\begin{figure}
\centering
\includegraphics[width=7cm, height=4cm]{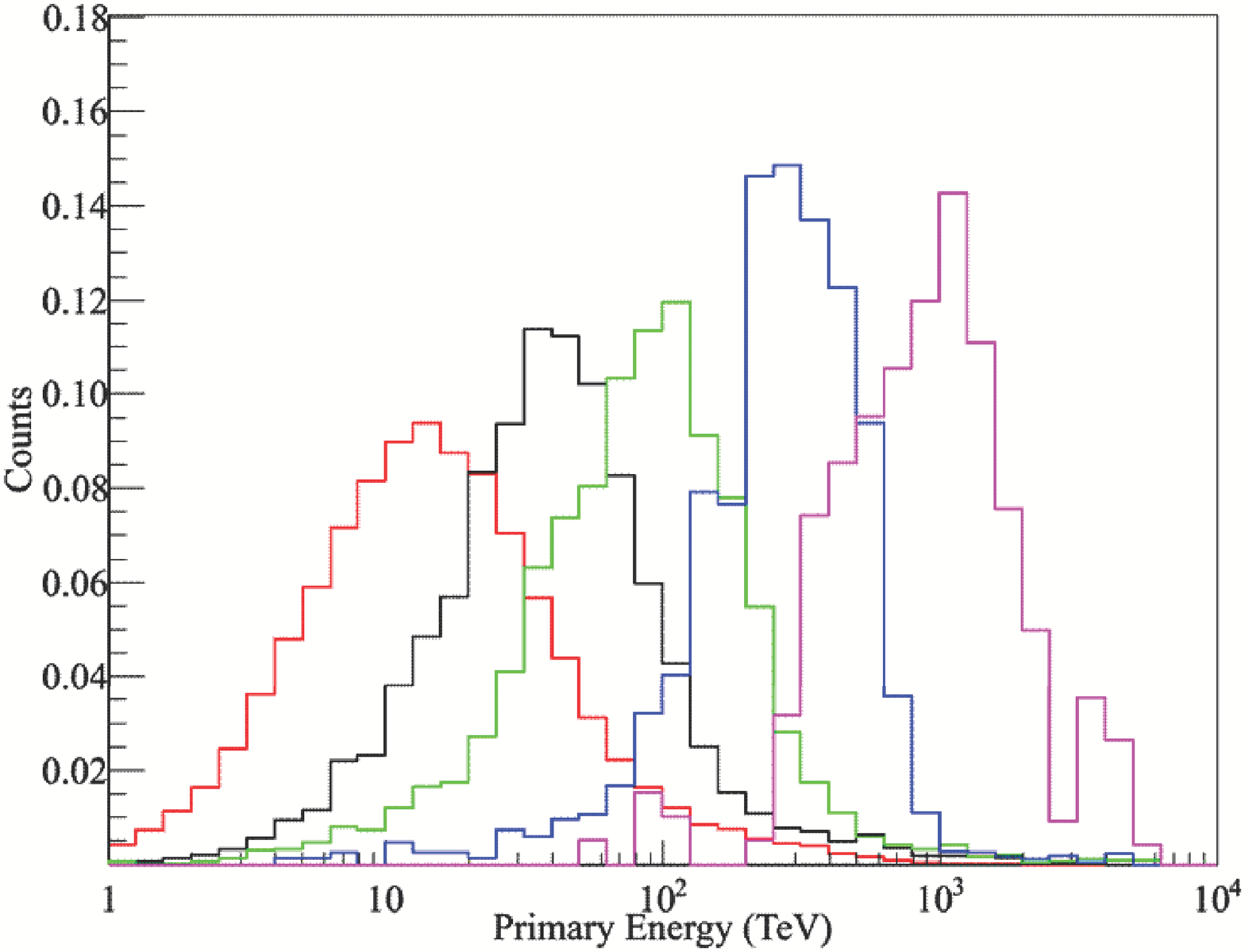}
\caption{\label{figEenrgyBand} Normalized event number as a function of the simulated primary energy in each of the five energy samples, based on MC data. The log-mean
energies of the five samples are 15 TeV (red), 50 TeV (black), 100
TeV (green), 300 TeV (blue), and 1 PeV (pink).}
\end{figure}

Figure~\ref{figEenrgyBand} shows the simulated primary energy
distributions of the five energy samples, as indicated by the dashed lines in Figure~\ref{figZenithEenrgy}. The uncertainty of the primary
energy estimate is dominated by the fluctuation of the AS. Event numbers in five energy samples are
$2.33\times10^{10}$ (15 TeV), $3.97\times10^{9}$ (50 TeV),
$1.96\times10^{9}$ (100 TeV), $2.71\times10^{8}$ (300 TeV), and
$5.72\times10^{7}$ (1 PeV), as listed in Table~\ref{tb1}.

\subsection{Analysis of the first harmonics of the anisotropy}
We employ the All-Distance Equi-Zenith Angle Method
\citep{2005ApJ...633.1005A,2006Sci...314..439A}, which has been
shown to be sensitive to probe the large-scale anisotropy, to
analyze the data. Details about this method can be found in
\citep{2005ApJ...633.1005A}. The sky of the horizontal coordinate is
divided into cells with a bin size of $1^{\circ}$ in both zenith (from
$0^{\circ}$ to $60^{\circ}$) and azimuth (from $0^{\circ}$ to
$360^{\circ}$). For the equatorial coordinate, the sky is divided
into cells of $2^{\circ}\times2^{\circ}$ between $0^{\circ}$ and
$360^{\circ}$ in the R.A. ($\alpha$) and between
$-30^{\circ}$ and $90^{\circ}$ in the decl. ($\delta$).
The two-dimensional (2D) map in the equatorial coordinate is then smoothed in a window, changing the window widths between 5 and 30$^{\circ}$.

To quantify the magnitude of the anisotropy, we project the
two-dimensional (2D) anisotropy map before smoothing onto the R.A. axis,
through averaging the relative intensities in all declinations from
$-30^{\circ}$ to $90^{\circ}$, to derive the one-dimensional (1D)
profile of the anisotropy. The R.A. is binned into 18
bins for this 1D analysis, and the 1D profile of the anisotropy is
fitted by the first order harmonic function in form of
\begin{equation}
R(\alpha)=1+A_{1} \cos(\alpha-\phi_{1}),
\label{eq1}
\end{equation}
where $R(\alpha)$ denotes the relative intensity of CRs at R.A. $\alpha$, $A_{1}$ is the amplitude of the first harmonics, and $\phi_{1}$ is the phase at which  $R(\alpha)$ reaches its maximum.

\section{Results}

\subsection{Sidereal Anisotropy map at 300 TeV}

\begin{figure*}[!htb]
\centering
\includegraphics[width=12cm, height=6cm]{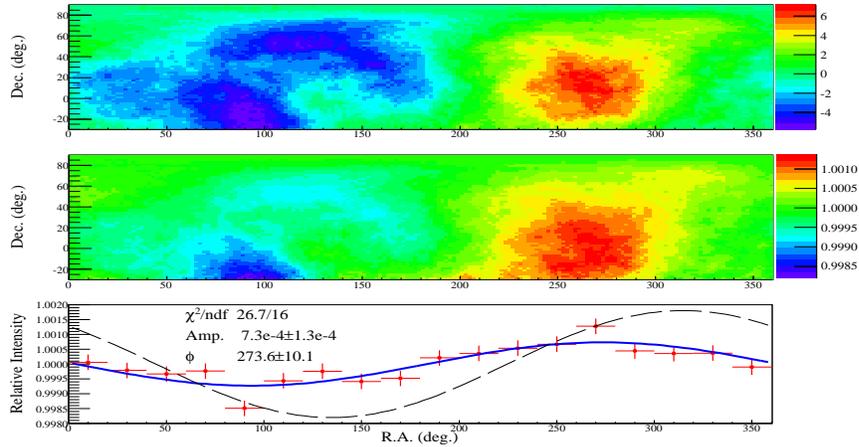}
\caption{\label{fig300TeVAnisotropy} Large-scale sidereal anisotropy
at 300 TeV by the Tibet AS Array. The 2D maps are smoothed with a
$30^{\circ}$ Gaussian kernel. The top and middle panels display the significance and relative intensity maps, respectively, while the bottom
one shows the 1D projection of the 2D map onto the R.A. axis. The blue curve shows the first harmonic fitting to
the data, and the black dashed line is the predicted Galactic CG
effect with an amplitude of $\sim0.19\%$.}
\end{figure*}

Figure~\ref{fig300TeVAnisotropy} shows the significance map and the
relative intensity map of the sidereal anisotropy  for the 300 TeV energy sample.
 A smoothing with an optimized window width of $30^{\circ}$ is applied in this figure.
 We combined the 300 TeV and 1 PeV samples together in this figure to increase the statistics. The total event number used in this figure is $3.28\times10^{8}$, and the median energy is
approximately 300 TeV.

From the significance map, we find that two regions are significant-that is,
an excess centered at ($\alpha=263^{\circ}$,
$\delta=11^{\circ}$) with a significance of $7.2\sigma$ and a deficit centered at ($\alpha=93^{\circ}$,
$\delta=-25^{\circ}$) with a significance of $-5.8\sigma$. Note that
the significance values are the pre-trial results. We conservatively
estimate a trial factor by assuming that all scans give
statistically independent results. Since the search for this excess
is performed over about 60$\times$180 cells, and 26 different
smoothing radii, the total trial factor is estimated to be about
$2.81\times10^{5}$. The post-trial significance of the excess is
$\sim5.2\sigma$. The deficit is no longer significant, fail to reach the
5$\sigma$ level, after the correction for the trials.

Because the acceptance of the detector decreases for larger zenith
angles, the relative intensity map is similar but not completely the
same as the significance map. An excess region centered at
($\alpha=269^{\circ}$, $\delta=-13^{\circ}$) with a maximum excess of +1.38$\times 10^{-3}$, and a deficit region centered at
($\alpha=87^{\circ}$, $\delta=-29^{\circ}$) with a maximum deficit of -1.80$\times 10^{-3}$ can be seen. Both the excess and
deficit regions are consistent with the results of IceCube at
400 TeV in the southern hemisphere \citep{2012ApJ...746...33A}. Combining these results gives us a
 full-sky picture of the sidereal anisotropy of GCRs at hundreds of
TeV.

The bottom panel in Figure~\ref{fig300TeVAnisotropy} also shows the 1D projection of the relative intensity
 before the smoothing onto the R.A. axis.
 The correlation among
different bins is carefully considered when calculating the
statistical errors and fitting the data with the harmonic function in equation \ref{eq1}.
If the correlation is not considered correctly, the errors
of the fitting parameters would be underestimated. The blue curve
shows the best-fitting result, with the fitting parameters indicated
in the figure. The significance of non-zero amplitude is
$5.6\sigma$, which shows that the obtained first harmonics are indeed
significant. The reduced $\chi^{2}$ value is 26.7/16, which means
that the first harmonic function can describe the 1D projected
profile well.

One of the possible origins of the sidereal anisotropy is the
Compton-Getting (CG) effect, due to the
orbital motion of the solar system around the Galactic center \citep{1935PhRv...47..817C}.
The relative intensity of this effect is carefully calculated by the MC method.  
Considering the location of the Tibet AS array ($90.522^{\circ}$E, $30.102^{\circ}$N), 
the velocity (220 km s$^{-1}$) of the orbital motion of the solar system around 
the Galactic center and the spectrum index (2.7) of the CRs energy spectrum, the intensity of the sky that is divided
into cells of $2^{\circ}\times2^{\circ}$ between $0^{\circ}$ and
$360^{\circ}$ in the R.A. ($\alpha$) and between
$-30^{\circ}$ and $90^{\circ}$ in the decl. ($\delta$) is calculated. Then the identical analyses are performed to this MC data sample.

The R($\alpha$) expected from this CG effect, shown as the black dashed line in Figure~\ref{fig300TeVAnisotropy},
has a maximum at ($\alpha=315^{\circ}$, $\delta=0^{\circ}$) and a minimum at ($\alpha=135^{\circ}$, $\delta=0^{\circ}$). Neither the
amplitude nor the phase of the large-scale anisotropy observed in
this work can be described in terms of the CG effect.

\subsection{Variation of CR sidereal anisotropy with the energy between 10 and 1000 TeV}
\begin{figure*}[!htb]
\centering
\includegraphics[width=12cm, height=8cm]{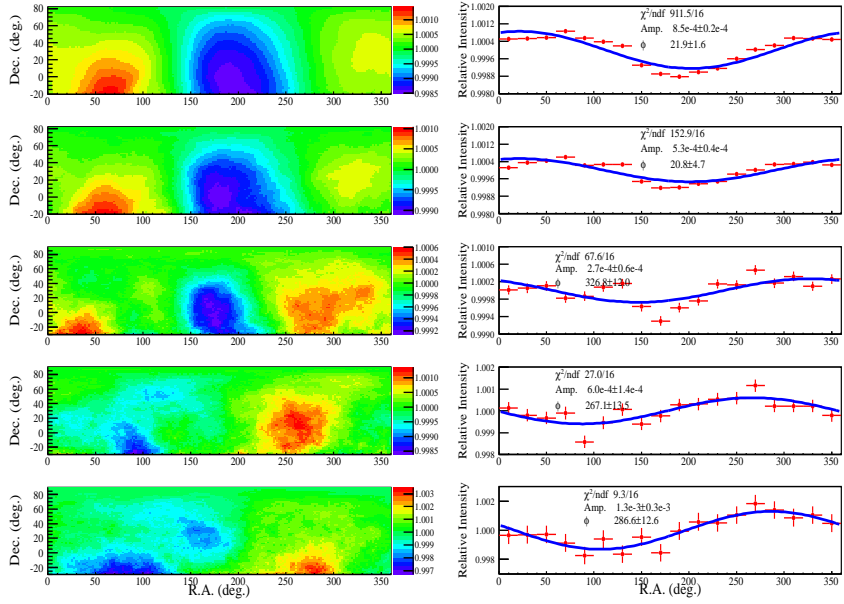}
\caption{\label{figTransition} 2D anisotropy maps in five energy samples (15, 50, 100, 300, and 1000 TeV, from top to bottom). Left
panels show the relative intensity maps (with $30^{\circ}$ smoothing), while
right panels show the 1D projections. The meaning of the blue curves in the right
panels is the same as in Figure \ref{fig300TeVAnisotropy}.}
\end{figure*}

Figure~\ref{figTransition} shows the variation of the sidereal
anisotropy with the energy between 10 TeV and 1 PeV. At 15 TeV and 50 TeV, the
tail-in and loss-cone features \citep{2006Sci...314..439A} are
observed with very high significances. An intensity excess in
the Cygnus region can also be seen. However, these features become
less significant above 100 TeV, being replaced with some new features. At
300 TeV and 1 PeV, the anisotropy maps are distinctly different from
those in 15-50 TeV. We can clearly see the phase of the 1D projection changing
with the primary energies, as seen in Table~\ref{tb1}, showing the best-fit parameters.

\begin{figure*}[!htb]
\centering
\includegraphics[width=12cm, height=8cm]{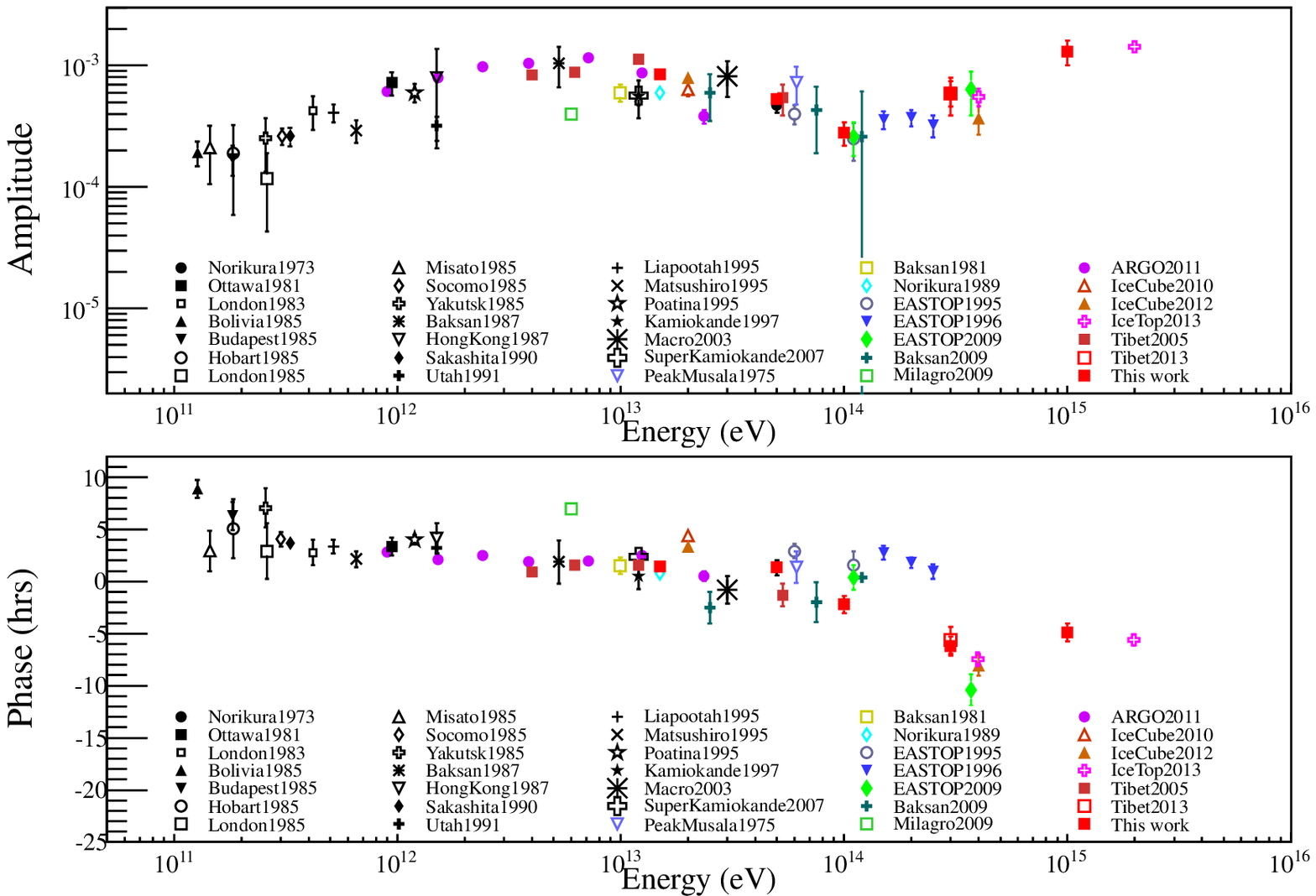}
\caption{\label{fig1DFit}The energy dependences of amplitude (top) and phase (bottom) of the first harmonics of the CRs anisotropy obtained in this work, and reported from previous measurements. They
are underground muon observations: Norikura(1973) \citep{1973ICRC....2.1058S}, Ottawa(1983) \citep{1981ICRC...10..246B}, London(1983)
\citep{1983ICRC....3..383T}, Bolivia(1985)
\citep{1985P&SS...33.1069S}, Budapest(1985)
\citep{1985P&SS...33.1069S}, Hobart(1985)
\citep{1985P&SS...33.1069S}, London(1985)
\citep{1985P&SS...33.1069S}, Misato(1985)
\citep{1985P&SS...33.1069S}, Socorro(1985)
\citep{1985P&SS...33.1069S}, Yakutsk(1985)
\citep{1985P&SS...33.1069S}, Banksan(1987)
\citep{1987ICRC....2...22A}, HongKong(1987)
\citep{1987ICRC....2...18L}, Sakashita(1990)
\citep{1990ICRC....6..361U}, Utah(1991)
\citep{1991ApJ...376..322C}, Liapootah(1995)
\citep{1995ICRC....4..639M}, Matsushiro(1995)
\citep{1995ICRC....4..648M}, Poatina(1995)
\citep{1995ICRC....4..635F}, Kamiokande(1997)
\citep{1997PhRvD..56...23M}, Marco(2003)
\citep{2003PhRvD..67d2002A}, SuperKamiokande(2007)
\citep{2007PhRvD..75f2003G}, and air shower array experiments:
PeakMusala(1975) \citep{1975ICRC....2..586G}, Baksan(1981)
\citep{1981ICRC....2..146A}, Norikura(1989)
\citep{1989NCimC..12..695N}, EASTOP(1995,1996,2009)
\citep{1995ICRC....2..800A,1996ApJ...470..501A,2009ApJ...692L.130A},
Baksan(2009) \citep{2009NuPhS.196..179A}, Milagro(2009)
\citep{2009ApJ...698.2121A},
IceCube(2010,2012) \citep{2010ApJ...718L.194A,2012ApJ...746...33A},
IceTop(2013) \citep{2013ApJ...765...55A}, ARGO-YBJ(2015) \citep{2015ApJ...809...90B}, Tibet(2005,2013)
 \citep{2005ApJ...626L..29A, ZFeng2013}.
}
\end{figure*}

Figure~\ref{fig1DFit} compares the amplitude and phase obtained in
this work with those reported so far from the deep underground muon experiments and
extensive AS experiments. Our results are in close agreement with other
results in similar energy regions in both the amplitude and the phase. It is interesting to note that a sharp increase of the
amplitude above 100 TeV can be seen
in the upper panel. The origins of this feature cannot be
explained with the conventional diffusion scenario of GCRs, and may provide us
with a new hint for understanding the origin and propagation of GCRs.

\subsection{Anisotropy in Solar Time and Antisidereal Time}
\begin{figure*}[!htb]
\centering
\includegraphics[width=12cm, height=8cm]{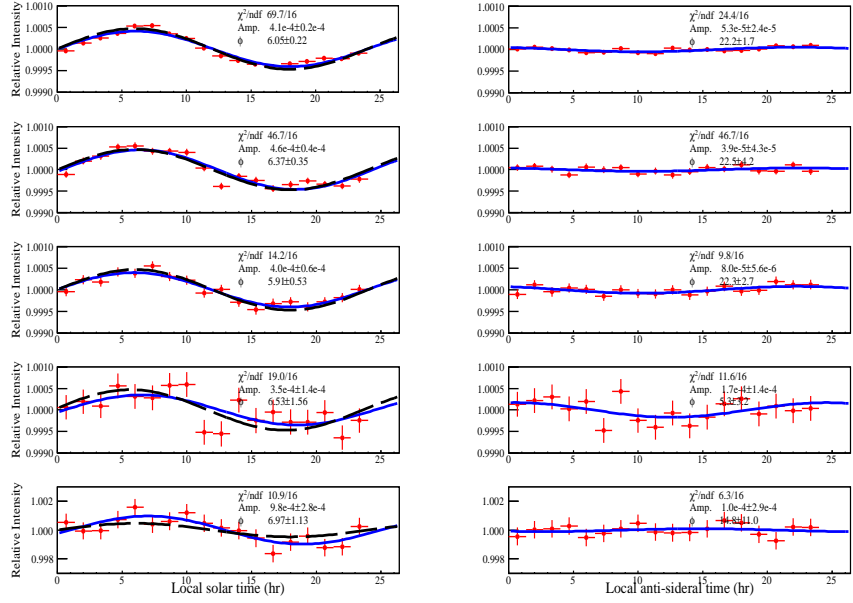}
\caption{\label{fig1DSolarTimeAntiSideralTime}
    Local solar time and antisidereal time daily variations measured by Tibet AS Array in five energy samples, 15, 50 , 100, 300, and 1000 TeV, from
top to bottom. Left panels show the solar time daily variations, with blue curves showing the first harmonic fit to the data
and the black dashed curves indicating the expected CG effect due to the Earth's orbital motion around the sun, with an
amplitude of 0.047\% and an phase of 6 hr. Right panels show the
antisidereal time daily variations and the first harmonic fitting
results. }
\end{figure*}

In order to confirm that the obtained anisotropy is not affected by the seasonal variation of the AS array performance, identical analyses are performed in the solar time and
antisidereal time frames in five energy samples. Figure~\ref{fig1DSolarTimeAntiSideralTime} shows the local solar time and antisidereal time daily variations measured by Tibet AS Array in five energy samples, and the best-fit parameters are also shown in Table~\ref{tb1}. The amplitude and phase in the solar time frame are in good agreement with the
expectation from the CG effect due to the terrestrial orbital motion
around the Sun ($A_{\rm sol,CG}=0.047\%$ and $\phi_{\rm sol,CG}=6.0$
hr). In all five energy samples, no significant anisotropy is
observed in the antisidereal time frame, ensuring that no
additional correction is required for the seasonal
effects. The observed results in the solar and antisidereal
time frames support the reliability of the observed sidereal anisotropy.

\section{Conclusion and Discussion}
Fifteen years data recorded by the Tibet AS array have been
analyzed to study the sidereal anisotropy of CRs. In this work, we improve the estimate of the primary CR energies
through a 2D cut in the $\sum\rho_{FT}-$sec$\theta$ plane, to explore the anisotropy including larger zenith angle events.
For the first time, we extend the analyzed decl. down to $-30^{\circ}$ to complete a full-sky coverage of
the anisotropy at hundreds of TeV energies by combining with
the IceCube's results at the South Pole. The 2D anisotropy map at $\sim300$ TeV obtained in this work is smoothly connected with IceCube's results at 400 TeV. The energy dependence of the large-scale sidereal anisotropy
 has been derived between 10 TeV and 1 PeV.
We measured the energy dependence of the first
harmonics of the anisotropy above 100 TeV, which may be
associated with local origins of GCRs.

The CG effect expected from the orbital motion of
the solar system around the Galatic center is not observed at 300 TeV, as shown in
Figure \ref{fig300TeVAnisotropy}. The basic picture that GCRs are
co-rotating with the local Galactic neighbors still holds
at this energy \citep{2006Sci...314..439A}. As pointed out earlier, the GCR rest frame may have a smaller
relative velocity with a different direction from neighboring stars and the
interstellar medium \citep{2012ApJ...746...33A}. This scenario is
possibly responsible for the GCR anisotropy observed at hundreds of TeV.

The strongest excesses at hundreds of TeV are from the direction of the
Galactic center, which may imply a Galactic center origin of GCRs at
these energies \citep{1367-2630-15-1-013053}. 
It is interesting to note that the highest-energy CR accelerators have been 
identified by the HESS telescope in the Galactic center \citep{2016Natur.531..476H}. However, the
energy dependences of the amplitude and phase cannot be easily
understood in a simple diffusion scenario with any types of GCR
sources.

The sharp increase of the amplitude above
100 TeV may imply an evolution of propagation parameters, such as spatial parameters
\citep{2015PhRvD..92h1301T,2016ApJ...819...54G}.
The knowledge of the
propagation of GCRs needs to be further improved for our full understanding
the properties of the anisotropy, especially in this high-energy
region where the conventional diffusion/drift models may not work any more.
Finally, we add to note that the measurements of the anisotropy above PeV, which is
possibly associated with the knee of GCRs, are very important to
advance our understanding of origin and propagation of GCRs.

\section{Acknowledgments}
The collaborative experiment of the Tibet Air Shower Arrays has been
performed under the auspices of the Ministry of Science and
Technology of China and the Ministry of Foreign Affairs of Japan.
This work was supported in part by a Grant-in-Aid for Scientific
Research on Priority Areas from the Ministry of Education, Culture,
Sports, Science and Technology, by Grants-in-Aid for Science
Research from the Japan Society for the Promotion of Science in
Japan, and by the Grants from the National Natural Science
Foundation of China and the Chinese Academy of Sciences. Zhaoyang Feng is supported by the Natural Sciences Foundation of China (Nos.11405182, Nos.1135010). C. Liu is supported by the Natural Sciences Foundation of China (Nos. 11405180).

\bibliographystyle{apj}
\bibliography{CRanisotropyEvolution10TeVPeV}

%\onecolumn
%\twocolumn

\begin{table*}[!htb]
\begin{center}
%\begin{deluxetable}{crrrrrrrrr}
%\tabletypesize{\scriptsize}
%\rotate
\caption{Fitting results of the first harmonic (Amplitude, Phase, and
Reduced $\chi^2$) in the sidereal (Columns 2-4), solar (Columns 5-7),
and Antisidereal (Columns 8-10) Times. The number of events in each energy
sample is given in Column 11. \label{tb1}}

%\tablewidth{0pt}
%\caption{

\begin{tabular}{ccccccccccc} \\ \hline

Energy & $A_{sid}$ & $\phi_{sid}$  & $\chi^{2}_{sid}/ndf$  & $A_{sol}$  & $\phi_{sol}$  & $\chi^{2}_{sol}/ndf$    & $A_{asid}$  & $\phi_{asid}$  & $\chi^{2}_{asid}/ndf$  & Number of Event \\
%E & $A_{sid}$ $\pm$ (stat.) & $\phi_{sid} \pm $(stat.) & $\chi^{2}_{sid}/ndf$  & $A_{sol}$ $\pm$ (stat.) & $\phi_{sol} \pm$ (stat.) & $\chi^{2}_{sol}/ndf$    & $A_{asid}$ $pm$ (stat.) & $\phi_{sol} \pm$ (stat.) & $\chi^{2}_{asid}/ndf$  \\
TeV & $10^{-4}$  & [$^{\circ}$] & & $10^{-4}$  & hr &  & $10^{-4}$  & hr  & &  \\ \hline
%}
%\startdata
15  &8.5$\pm$0.2  & 21.9$\pm$1.6   & 911./16  &  4.1$\pm$0.2  &  6.05$\pm$0.22  & 69.7/16. & 0.53$\pm$0.24  & 22.2$\pm$1.7 & 24.4/16 &  $2.33\times10^{10}$\\
50   &5.3$\pm$0.4  & 20.8$\pm$4.7   &152.9/16  &  4.6$\pm$0.4  &  6.37$\pm$0.35  & 46.7/16 & 0.39$\pm$0.43  & 22.5$\pm$4.2 &46.7/16 & $3.97\times10^{9}$\\
100  &2.7$\pm$0.6  & 326.8$\pm$12.0   &67.6/16  &  4.0$\pm$0.6  &  5.91$\pm$0.53  & 14.2/16. & 0.80$\pm$0.56  & 22.3$\pm$2.7 &9.8/16 & $1.96\times10^{9}$\\
300  &6.0$\pm$1.4  & 267.1$\pm$13.5   & 27.0/16. & 3.5$\pm$1.4  & 6.53$\pm$1.56  & 19.0/16. & 1.7$\pm$1.4  & 5.3$\pm$3.2 &11.6/16 & $2.71\times10^{8}$\\
1000  & 13.0 $\pm$3.0  & 286.6$\pm$12.6   & 9.3/16 &  9.8$\pm$2.8  & 6.97$\pm$1.13  & 10.9/16. & 1.0$\pm$2.9  & 14.8$\pm$11.0 &6.3/16 & $5.72\times10^{7}$\\

\hline
%\enddata
\end{tabular}
\end{center}

\end{table*}
%\end{deluxetable}

\end{document}